\documentclass[reprint, prd, twocolumn,floatfix]{revtex4}
\usepackage{hyperref}
\usepackage[english]{babel}
\usepackage[latin1]{inputenc}
\usepackage{enumerate}
\usepackage{amsfonts}
\usepackage{amssymb}
\usepackage{amsmath}
\usepackage{dcolumn}
\usepackage{bm}
\usepackage{graphicx, graphics}
\usepackage{graphicx}
	\begin{document}
		\title{Decay constants in AdS/QCD: a different approach}
	
	\author{Miguel Angel Martin Contreras}
\email{miguelangel.martin@uv.cl}
\affiliation{%
 Instituto de F\'isica y Astronom\'ia, \\
 Universidad de Valpara\'iso,\\
 A. Gran Breta\~na 1111, Valpara\'iso, Chile
}

\author{Alfredo Vega}%
 \email{alfredo.vega@uv.cl}
\affiliation{%
 Instituto de F\'isica y Astronom\'ia, \\
 Universidad de Valpara\'iso,\\
 A. Gran Breta\~na 1111, Valpara\'iso, Chile
}
		\begin{abstract}
The holographic recipe for the calculation of decay constants is revisited. Starting from the holographic 2-point function and using the fact that normalizable bulk modes scale as $z^{\Delta-S}$, with $S$ the spin, we can obtain a consistent expression that depends on the value of the mode at the boundary, not the derivative. We apply our decay constant expression to other AdS/QCD (static and dynamic) models proving its consistency. We also demonstrated that our approach is equivalent to the usual holographic prescription. 
\end{abstract}

\maketitle
\section{\label{sec:level1}Introduction}
One of the most interesting and prolific applications of the AdS/CFT correspondence \cite{Maldacena:1997re,Aharony:1999ti} is the holographic description of QCD, that can be done with two possible methodologies: top-down or bottom-up models. In the former, we tried to fix the dual QFT theory with the QCD \cite{Karch:2002sh,Erdmenger:2007cm}. In the later we do the opposite: we tried to fix the gravity background to describe the QCD phenomenology.  

For example, confinement \emph{a la} top-down is realized with the interaction of open strings attached to the conformal boundary with the geometric background. In the bottom-up case, confinement is realized by deforming the AdS space. This deformation is manifested in a holographic potential constructed from the geometry itself and other background fields, usually the dilaton. Hadrons appear here as bounded states of this potential. In this paper, we will focus on bottom-up approaches.

Altogether with the meson spectra, with the bottom-up models is possible to explore other hadronic observables as the form factors (e.g., see \cite{Brodsky:2007hb, Abidin:2009aj, Abidin:2009hr, Brodsky:2011xx, Gutsche:2012bp, Gutsche:2015xva, Ballon-Bayona:2017bwk}), structure functions (e.g., see \cite{Polchinski:2002jw, BallonBayona:2007rs,Cornalba:2008sp, Pire:2008zf,Braga:2011wa}), decay constants (e.g., see \cite{Grigoryan:2007my, Ballon-Bayona:2014oma, Braga:2015jca, Braga:2017bml}), etc. In this work we will focus on the decay constants for scalar and vector mesons written in AdS-like QCD models, such as the soft wall model. Other approximation to the decay constant calculation can be found in \cite{Braga:2015jca}, but in the context of heavy quarkonium.

The calculation process of the decay constant, as it was exposed in \cite{Karch:2006pv}, requires to compute the second derivative of the eigenmode associated to the bulk field, dual to the meson, at the boundary located at $z\to 0$. Unless the solution is analytical, in the numerical process, the decay result is strongly tied to the value of the UV cutoff used to describe the boundary numerically. For example, in the scalar meson case, the 2-point function, and consequently, the decay constant, has a $z^{-3}$ factor coming from the geometry, that makes unstable any result computed: small changes in the numerical tolerance are translated into bigger numerical errors in the holographic decay constant.  Inspired by this fact we tried to find an alternative method to compute decay constants, allowing us to avoid this numerical issue. The basis of the idea presented here has its grounds on the behavior of the eigenmodes at the boundary. According to the AdS/CFT correspondence, bulk field should scale as $z^{\Delta-S}$ when $z\to 0$, with $\Delta$ being the conformal dimension. In AdS-like backgrounds, this specific form of the solutions makes it possible to re-write the derivatives near to the boundary as the eigenmode itself evaluated at the same place. All of the divergent contributions are naturally suppressed, as the eqn. \eqref{decay-const-2} demonstrates. Thus, as a direct conclusion, the holographic decay constants depend only on three aspects: the value of the dilaton field at the boundary, the normalization constant coming from the bulk action and the value of the eigenmode solution at the origin. This reduces the numerical error due to the derivative calculations.   

This document is organized as follows. In section \ref{geom} we define the geometric configuration and the action for the bulk fields dual to mesons. In section \ref{Holo-decay} we do a summary of the ideas of the holographic calculation of the decay constants starting from the holographic 2-point function. We also present there our main result for the decay constants in terms of the bulk modes evaluated at the boundary. Section \ref{test} is devoted entirely to test our formula for the decays with some of the AdS/QCD models available in the literature. Finally, section \ref{conclu} collects the conclusions and perspectives of this work. 
 
\section{Geometric configuration} \label{geom}
To describe the holographic frame to construct the decay constants, we will focus on a 5-dimensional AdS-like geometry given by the line element

\begin{equation}
dS^2=e^{2\,A(z)}\left[dz^2+\eta_{\mu\nu}\,dx^\mu\,dx^\nu\right],    
\end{equation}

\noindent where $\eta_{\mu \nu}$ is the Minkowski 4-dimensional metric tensor and $A(z)$ is the warp factor, that can be fixed to be the usual Poincare one: $A(z)=\log R/z$.

Scalar and vector mesons are described in AdS/QCD models by the bulk fields. The normalizable part of such fields is dual the hadronic states, while the non-normalizable part is connected to the operators that create mesons. In this case, they are dual to the electromagnetic currents of the form $J_\mu=e\,\bar{\psi}\,\gamma_\mu\,\psi$. These bulk fields are defined, as usual, by the 5-dimensional action

\begin{equation}\label{action-main}
I_\text{Mesons}=I_\text{Vector}+I_\text{Scalar},    
\end{equation}

\noindent where

\begin{eqnarray*}
I_\text{Vector}&=&-\frac{1}{4\,g_V^2}\int{d^5x\,\sqrt{-g}\,e^{-\Phi(z)}\,F_{mn}\,F^{mn}},\\
I_\text{Scalar}&=&\frac{1}{2\,g_S^2}\int{d^5x\,\sqrt{-g}\,e^{-\Phi(z)}\,\left[\partial_m\,S\,\partial^m\,S+M_5^2\,S^2\right]},
\end{eqnarray*}

\noindent with $A_m(z,x)$ and $S(z,x)$ are the bulk vector and bulk scalar fields respectively. The constants $g_V$ and $g_S$ fix the action units and also contribute to the value of the holographic decay constants. These constants are fixed by the large $N_c$--QCD 2-point function at the large $q^2$ limit \cite{Hong:2004sa}. The bulk mass associated with the fields defines the identity of the hadronic state: for vector, mesons is zero, for scalar mesons takes the value of $-3$. 

Notice also that the action \ref{action-main} can be extended to AdS/QCD models with dynamical backgrounds. This will be done in section \ref{test}.

\section{Holographic
construction of the decay constants in a nutshell}\label{Holo-decay}

In general, in any AdS/QCD model, decay constants for mesonic states are defined from the 2-point function \cite{Karch:2006pv,Erlich:2005qh,Grigoryan:2007my,Afonin:2010fr,Colangelo:2008us,Braga:2015jca}. This object is holographically constructed from the on-shell boundary action. The generic result obtained for the 2-point function is of the form

\begin{equation}\label{2-point-f}
\Pi\left(-q^2\right)=-\left.\frac{e^{-B\left(z\right)}}{\mathcal{K}\,(-q^2)}\,\partial_z\,V\left(z,q\right)\right|_{z\to0},    
\end{equation}

\noindent where $B(z)=\Phi(z)+\beta\,A(z)$ and $\mathcal{K}$ is a normalization constant fixed with large $Q^2\equiv-q^2$ behavior of the large $N_c$ QCD 2-point function. The parameter $\beta$ carries the information of the spin: in case of scalar mesons $\beta=-3$ and for vector mesons $\beta=-1$.  In general, $\beta=-(3-2S)$ for bosonic sector. 

The bulk to boundary propagator $V(z)$ comes from the equations of motion for the non-normalizable part of the bulk field associated with mesons. As it was shown by \cite{Grigoryan:2007vg,Hong:2004sa}, we can write $V(z)$ in terms of the set of eigenfunctions $\psi_n(z,q)$ defined by the normalizable part of the bulk field. This normalizable part is dual to the mesonic modes and gives rise to the mass spectrum, understood it as the eigenvalue spectrum $M_n^2$. Both sets are obtained  from the  Sturm-Liouville form of the equations of motion for  such bulk fields:

\begin{equation}\label{Sturm-Liouvile}
\partial_z\left[e^{-B}\,\partial_z\,\psi\right]+M_n^2\,e^{-B}\,\psi-M_5^2\,e^{2A-B}\,\psi=0.    
\end{equation}

This equation defines a boundary value problem (BVP) with the following set of boundary conditions

\begin{eqnarray}
\left.\psi(z,q)\right|_{z\to0}&=&z^{\Delta-S},\\
\left.\psi(z,q)\right|_{z\to \infty}&=&0.
\end{eqnarray}

Recall that $M_5^2$ defines the \emph{hadronic identity} of the bulk-field via the relation with the conformal dimension\cite{Aharony:1999ti}

\begin{equation}\label{bulk-mass}
M_5^2\,R^2=(\Delta-S)(\Delta+S-4),	    
\end{equation}

\noindent where for mesons $\Delta=3$.

Finally, we can introduce the Green's function associated with the BVP defined by the equation \eqref{Sturm-Liouvile} as

\begin{equation}\label{Green-F}
   G(z,z',q)=\sum_{n=1}^\infty\frac{\psi(z,q)\,\psi(z',q)}{-q^2-M_n^2+i\,\epsilon}.   
\end{equation}

From the second Green's identity, we can construct the connection between the bulk-to-boundary propagator $V(z,q)$ with the Green's function as

\begin{equation}\label{eq:bulk-Green-F}
V(z,q)=\lim_{z\to 0}e^{-\phi(z)+A(z)}\,\partial_z\,G(z,z';q).    
\end{equation}

Now we can insert expressions \eqref{eq:bulk-Green-F} and \eqref{Green-F} into the 2-point function to get

\begin{multline}
\Pi(-q^2)=\frac{1}{\mathcal{K}\,M_n^2}\,\lim_{z,z'\to0}e^{-2\,\Phi(z)}\,e^{-(\beta-1)A(z)}\,\times\\
\times\sum_{n=1}^{\infty}{\frac{\partial_z\,\psi_n(z,q)\,\partial_{z'}\,\psi_n(z',q)}{-q^2-M_n^2+i\epsilon}}.\label{2-point-function-S}     
\end{multline}

Evaluating the limit we can write the standard form in terms of second derivatives of the eigenfunction $\psi(z,q)$:

\begin{equation}\label{2-point-holo}
\Pi(-q^2)=\frac{1}{\mathcal{K}\,M_n^2}\lim_{z\to0}\,e^{-2\,\Phi(z)}e^{-(\beta+1)A(z)}\sum_{n=1}^{\infty}\frac{\left|\psi''_n(z,q)\right|^2}{-q^2+M_n^2+i\,\epsilon}.  
\end{equation}

The 2-point function is a spectral decomposition of all the states present in a radial Regge trajectory, that has a general form given by 

\begin{equation}\label{2-point-exp}
    \Pi(-q^2)=\sum_{n=1}^{\infty}\frac{f_n^2}{-q^2+m_n^2+i\epsilon},
\end{equation}

\noindent where $f_n$ are the decay constants, given in units of energy squared. 

Comparing \eqref{2-point-holo} with \eqref{2-point-exp}, we can conclude that the decay constants are 

\begin{equation}\label{decay-cons-1}
F_n^2=f_n^2\,M_n^2=\frac{1}{\mathcal{K}}\,\lim_{z\to0}{\,e^{-2\,\Phi(z)}\,e^{-(\beta+1)\,A(z)}\left|\psi_n''(z,q)\right|^2}.
\end{equation}

It is customary to introduce the decay constant as $F_n=M_n\,f_n$, where $F_n$ is defined by $\left.\langle0\right|J_\mu^a\left|M^b\rangle\right.=F_n\,\delta^{a\,b}\,\epsilon_mu$ for a given meson $M$. See for example chapter 2 of \cite{Manohar2000}. 

The expression \eqref{decay-cons-1} is just a generalization of the expression given by \cite{Karch:2006pv,Erlich:2005qh,Colangelo:2008us} for vector and scalar mesons in the context of the soft wall and hard wall models. It also has the advantage that it can be adapted to any other AdS/QCD model. Only need to specify the dilaton, the warp factor and obtain the solutions. As a test, we will compute the decay constants for scalar and vector mesons in the soft wall model.


\subsection{Ansatz for the eigenfunctions}

According to the holographic prescription, the normalizable solutions should behave as $z^{\Delta-S}$. This fact motivates us to propose the power expansion of the eigensolutions at the conformal boundary as 

\begin{equation}\label{anzats}
\psi_n(z,q)=\mathcal{C}_{\Delta-S}(q,n,S)\,z^{\Delta-S}+\sum_{m>\Delta-S}^\infty{\mathcal{C}_m(q,n,S)\,z^m},
\end{equation}

\noindent where $\mathcal{C}_{\Delta-S}(q,n,S)$ is the lowest coefficient in the power expansion. This coefficient is different from the normalization  constant, constructed as

\begin{equation}
    \int{dz\,e^{-B(z)}\,\psi_n(z,q)\,\psi^*_m(z,q)}=\delta_{n\,m}.
\end{equation}



The second derivative, in the limit $z\to0$ can be written in terms of the eigenfunction $\psi_n$ itself as 

\begin{eqnarray}\notag
\left.\psi_n''(z,q)\right|_{z\to 0}&=&(\Delta-S)(\Delta-S-1)\,\mathcal{C}_{\Delta-S}(q,n,S)\,z^{\Delta-S-2}\\ \label{derivative-limit}
&=&(\Delta-S)(\Delta-S-1)\left.\frac{\psi_n(z,q)}{z^{2}}\right|_{z\to 0},
\end{eqnarray}

\noindent where  the constant $\mathcal{C}_{\Delta-S}$ extracted from \eqref{anzats}.

In limit $z\to 0$, all the terms, with a generic $z^{m-\Delta+S}$ dependence vanish since $m>\Delta-S$ by construction. 

Therefore, the decay constants can be written as 

\begin{equation}\label{decay-cons-2}
    F_n^2=\frac{1}{\tilde{\mathcal{K}}}\,\lim_{z\to0}{\,e^{-2\,\Phi(z)}\,e^{-(\beta+1)\,A(z)}\,\left|\frac{\psi_n(z,q)}{z^2}\right|^2},
\end{equation}

\noindent with $\tilde{\mathcal{K}}=\mathcal{K}/(\Delta-S)^2(\Delta-S-1)^2$. 

This expression connects decay constants with the AdS normalizable mode directly, in a similar form as the Van Royen-Weisskopf formula \cite{VanRoyen1967} connects the decay constants with the mesonic wave function at the origin. 

In the case of asymptotic AdS spaces constructed dynamically, the second derivative written in \eqref{2-point-holo} cannot be constructed directly. Thus, it is customary to use the first derivative definition of the decay, i.e.,

\begin{equation}
F_n^2=\frac{1}{\mathcal{K}}\lim_{z\to0}\,e^{-2\,\Phi(z)}e^{-(\beta-1)A(z))}\left|\psi_n'(z,q)\right|^2,    
\end{equation}

\noindent that can be read off directly from holographic 2-point function. Following the same procedure as we did above, we can prove that 

\begin{equation}\label{decay-cons-3}
F_n^2=\frac{(\Delta-S)^2}{\mathcal{K}}\,\lim_{z\to0}\,e^{-2\,\Phi(z)-(\beta-1)A(z)}\left|\frac{\psi_n(z,q)}{z}\right|^2.    
\end{equation}

We will prove this expression in the case of dynamical models in the next section. 

\subsection{Simplified form of the decay constants}
If we go further with our analysis of the decay constant expression, we can use the ansatz \eqref{anzats} into \eqref{decay-cons-2} considering an AdS spacetime defined by the usual Poincare patch, i.e., $A(z)=\log(R/z)$. Therefore, the decay constants are written as 

\begin{multline}
  F_n^2=\frac{\mathcal{C}^2_{\Delta-S}(n,q,S)(\Delta-S)^2(\Delta-S-1)^2}{\mathcal{K}}\,\\
  \lim_{z\to 0}\,e^{-2\,\Phi(z)}\,\left(\frac{R}{z}\right)^{-(\beta+1)}\,\left[z^{2\Delta-2S-4}+\mathcal{O}\left(z^{2m-4}\right)\right].
\end{multline}

The limit can be evaluated as follows

\begin{multline}
\lim_{z\to 0}\,e^{-2\,\Phi(z)}\,\left(\frac{R}{z}\right)^{-(\beta+1)}\,\left[z^{2\Delta-2S-4}+\mathcal{O}\left(z^{2m-4}\right)\right]=\\
=R^{-(\beta+1)} \lim_{z\to 0}\,e^{-2\,\Phi(z)}\,\left[z^{2\Delta-6}+\mathcal{O}\left(z^{2m+\beta+3}\right)\right]. 
\end{multline}

For mesons  $\Delta=3$, therefore the first term of the power series in $z$ is always $z^0$. The next orders in the expansion are $2m>\left|\beta+3\right|>0$. Thus, in the limit $z\to 0$, all of them vanish. Therefore, the decay constants are finally written as

\begin{equation}\label{decay-const-2}
F_n^2=\frac{R^{-(\beta+1)}e^{-2\,\Phi_0}}{\mathcal{K}}\,\mathcal{C}^2_{\Delta-S}(n,q,S)(\Delta-S)^2(\Delta-S-1)^2, 
\end{equation}

\noindent where $\Phi_0$ means the value of the dilaton at the conformal boundary. If we start from the expression \eqref{decay-cons-3} instead of \eqref{decay-cons-2}, it is straightforward to prove that we arrive at a similar expression for the decay constants. 

Thus, we conclude that the decay constant depends on the first coefficient on the power series only, as we would expect from expression \eqref{decay-cons-2}. This is a powerful form of testing any AdS/QCD model easily, altogether with the mass spectrum.

\section{Testing the new expression} \label{test}
In the following paragraphs, we will prove our expressions \eqref{decay-const-2} for some specific models in the AdS/QCD literature. 

\subsection{Hard wall model}
Hard wall model was introduced in \cite{BoschiFilho:2002ta,Polchinski:2002jw}. It considers the appearance of confinement as a consequence of putting a cutoff  (hard wall) at some finite position $z_\text{HW}$ of the AdS space. This wall is endowed with an extra Dirichlet boundary condition. In other (geometrical) words, this wall can be considered as a D-brane. The hard wall makes possible the emerging of eigenmodes dual to hadrons at the conformal boundary.

In this model, a $B$ function is defined in terms of the warp factor only, so the dilaton is fixed to be zero in all of our expressions. Following this prescription, we can find the solution of the equation of motion \eqref{Sturm-Liouvile} in terms of Bessel functions of first kind $J_1(x)$ as 

\begin{equation}
\psi_{n,S}(z,q)=\frac{\sqrt{2}\,z^{\frac{1-\beta}{2}}}{z_\text{HW}\,\left|J_2(M_n\,z_\text{HW})\right|}J_1(M_n\,z
),    
\end{equation}

\noindent where the mass spectrum $M_n$ is given by the zeroes $\alpha_{1,n}$ of the Bessel function, i.e.,  $J_1(\alpha_{1,n})=0$, as follows:
\begin{equation}
M_n=  \frac{\alpha_{1,n}}{z_\text{HW}}.  
\end{equation}

With these definitions, we can calculate the decay constants. To do so, we follow the expression \eqref{decay-cons-2}:

\begin{equation}
\left.\frac{\psi_{n,S}(z,q)}{z^{\Delta-S}}\right|_{z\to0}=\frac{M_n}{\sqrt{2}\,z_\text{HW}\,\left|J_2(M_n\,z_\text{HW})\right|}.    
\end{equation}

Thus, we can write the decay constants in the hard wall model as

\begin{equation}
F_n^2=\frac{(\Delta-S)^2(\Delta-S-1)^2}{\mathcal{K}}\,\frac{M_n^2}{2\,z_\text{HW}^2\,\left|J_2(\alpha_{1,n}\,z_\text{HW})\right|^2},
\end{equation}

\noindent that  evaluated  in the scalar and vector mesons case gives:

\begin{eqnarray}
F_{n,\text{Vector}}^2&=&\frac{2\,M_n^2}{\mathcal{K}\,z_\text{HW}^2\,\left|J_2(\alpha_{1,n}\,z_\text{HW})\right|^2}\\
F_{n,\text{Scalar}}^2&=&\frac{18\,M_n^2}{\mathcal{K}\,z_\text{HW}^2\,\left|J_2(\alpha_{1,n}\,z_\text{HW})\right|^2}.
\end{eqnarray}

These expressions are consistent with those exposed in \cite{Grigoryan:2007my}, but they are, in fact, quite different. The fundamental difference raises in the origin of the expressions. That work starts from the bulk-to-boundary propagator itself, which is written in terms of $J_0(x)$ Bessel functions. In our case, we start from the eigenmodes that are, at is expected, different: the former scales as $z^{-\Delta+S}$, while the other does as $z^{\Delta-S}$.

To see how both expressions reproduce the same physics, we can study the high excitation number limit. In this case, $M_n=\pi\,n\,/ z_\Lambda=\pi\,n\,\Lambda$, with $\Lambda$ some energy scale associated to the theory \cite{Polchinski:2002jw}. Under these conditions, the vector meson decay constant can be written as

\begin{equation}\label{decay-large-n}
F_n=\frac{\sqrt{2}\,M_n}{\sqrt{\mathcal{K}}\,z_\text{HW}\,\left|J_2(\alpha_{1,n})\right|} \propto \Lambda^2\,n^{3/2}\,N_c,
\end{equation}

\noindent where we have used the fact that $\mathcal{K}$ should scale with $N_c^2$, according to the matching condition at high $q^2$ in the large $N_c$ limit. Notice that expression \eqref{decay-large-n} has the supergravity limit, $F_n\propto n^{3/2}$, expected for conserved currents in this kind of dual models \cite{Hong:2004sa}. The same behavior can be observed for the scalar case. Also, notice this scaling behavior in inherit by the ``conformal nature" that the hard wall has. On the other hand, this scaling behavior is also expected in the large-$N_c$ QCD phenomenology \cite{Afonin:2010fr}, since $F_n\sim M_n^2\,\partial_n\,M_n^2$ (See \cite{Afonin:2004yb} for details).

If we modify the conformal properties of the model by introducing dilaton fields, this limit does not hold anymore.  

\subsection{Soft wall model}

The soft wall model was introduced in \cite{Karch:2006pv}, and it has provided a lot of successful approximations to many QCD phenomena, as structure functions, form factors, and meson spectra.

In general, the soft wall model is defined by a quadratic static dilaton field living in a geometry described by the Poincare patch. This is expressed in the $B$ function as

\begin{equation}
B(z)=\kappa^2\,z^2+\beta\,\log\left(\frac{R}{z}\right),    
\end{equation}

\noindent where $\kappa$ is the energy scale used to fix the radial Regge linear trajectory. Using the equation \eqref{Sturm-Liouvile} and the definition of the bulk mass  \eqref{bulk-mass} we can arrive at the general form of the solutions for mesons in the soft wall model in terms of associated Laguerre polynomials $L_n^a(x)$ as:

\begin{equation}
\psi_{n,S}(z,q)=\sqrt{\frac{2}{n+1}}\,\kappa^2\,z^{3-S}\,L_n^1(\kappa^2\,z^2).
\end{equation}

In order to apply \eqref{decay-const-2} we need to compute $\mathcal{C}_{\Delta-S}$. To do this, we can use the expression \eqref{derivative-limit} and the properties of the Associated Laguerre polynomials to get:

\begin{equation}
 \mathcal{C}_{\Delta-S}=  \left.\frac{\psi_n(z,q)}{z^{\Delta-S}}\right|_{z\to 0}= \sqrt{2(n+1)}\,\kappa^2.
\end{equation}

Thus the decay constants are

\begin{equation}
 F_n^2=\frac{2\,\kappa^4\,R^{-(\beta+1)}}{\mathcal{K}}\,\left(n+1\right)(3-S)^2(2-S)^2.
\end{equation}

Evaluating for scalar ($\beta=-3$) and vector ($\beta=-1$) separately, we arrive to

\begin{eqnarray}\label{decay-swm-vec}
F^2_{\text{vector},n}&=&\frac{8\,\kappa^4\,(n+1)}{\mathcal{K}},\\ \label{decay-swm-scr}
F^2_{\text{scalar},n}&=&\frac{72\,\kappa^4\,(n+1)}{\mathcal{K}}.
\end{eqnarray}

Notice that, beyond the normalization $\mathcal{K}$ fixed by examining the large $q^2$ behavior of the 2-point functions, the expressions are equivalent to those obtained for vector mesons [eqn. (17) in \cite{Karch:2006pv} with $\mathcal{K}=g_5^2$] and scalar mesons [eqn. (33) in \cite{Colangelo:2008us} with $\mathcal{K}\to 9\,k\,R/2$]. Recall that these scales are fixed by the high $q^2$ limit at large $N_c$ QCD \cite{Erlich:2005qh,Afonin:2010fr}.

\begin{figure}[ht]
\includegraphics[width= 3.5 in]{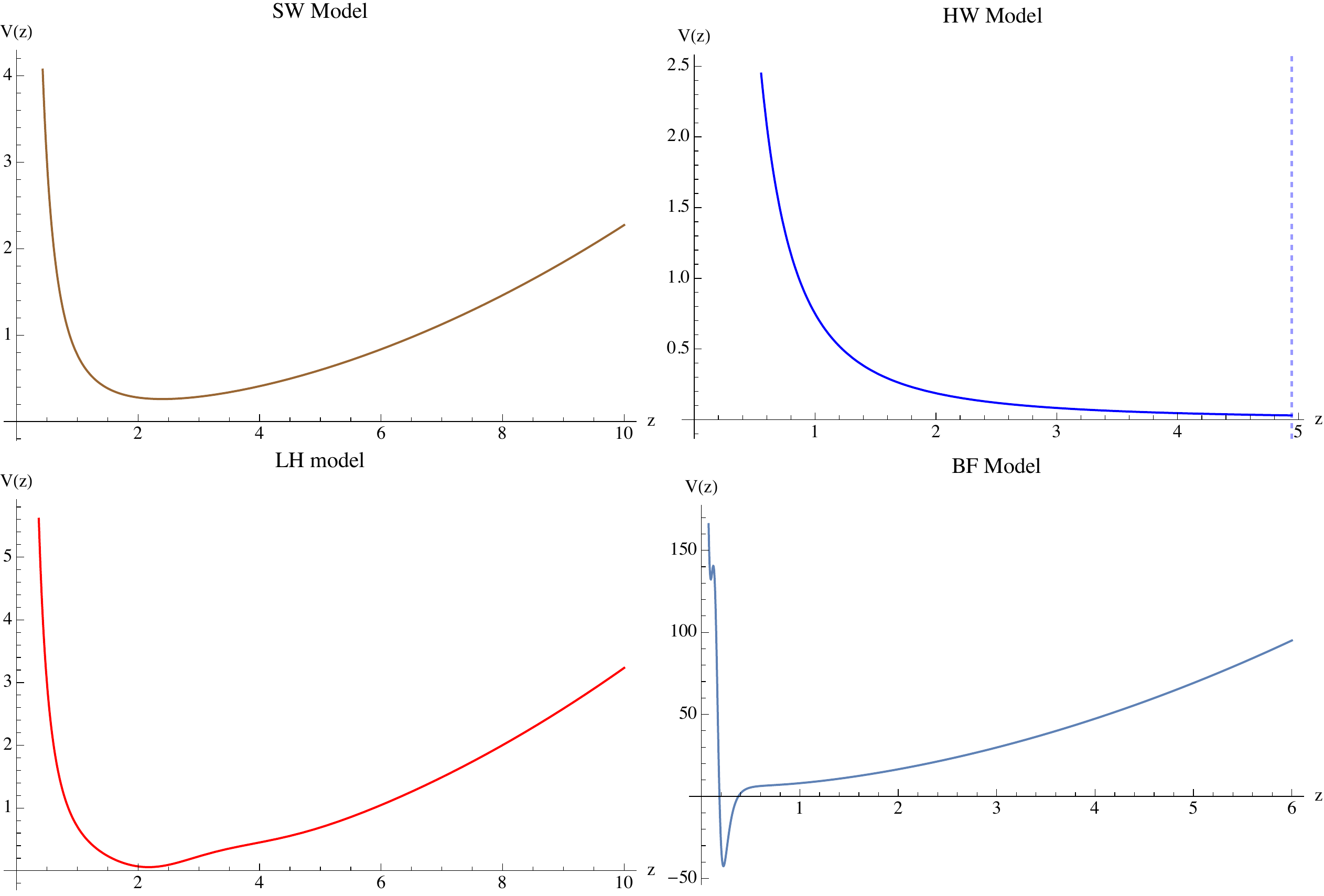}
\caption{\label{fig:one} This panel depicts the holographic potentials associated to discussed models in this work. The plots were made using the parameters given by the original authors, except for the hard wall model, where we used the $\rho(770)$ mass to fix the wall locus. All of the potentials correspond to vector meson cases.}
\end{figure}

\subsection{Braga and Ferreira Dilaton}

In the manuscript \cite{Braga:2019yeh}, authors discuss a deformed version of static quadratic soft wall model dilaton to address masses and decay constants for the heavy vector quarkonium. The proposed dilaton has the form:

\begin{equation}
\Phi(z)=k^2\,z^2+M\,z+\tanh\left(\frac{1}{M\,z}-\frac{k}{\sqrt{\Gamma}}\right),    
\end{equation}

\noindent where the parameters are defined as follows: $k$ is associated to the quark mass, $\Gamma$ is connected to the string tension present in the strong interaction between quark inside the meson; and $M$ is the mass scale associated to the decay process of the vector mesons going into leptons, i.e, $V_n\to l\bar{l}$.  In other words, $M$ gives the correct energy scale for the annihilation of any vector meson state, represented by the decay constant:   
$\langle0\left|J_\mu(0)\right|V_n\rangle\sim F_n$.  

Using this dilaton with geometrical frame given by the usual Poincare patch, and following the same procedure as we did with the other examples,  we can write a Sturm-Liouville equation as: 

\begin{multline}
-\psi''(z)-\left[\frac{1}{M\,z^2}\,\text{sech}^2\left(\frac{1}{M\,z}-\frac{k}{\sqrt{\Gamma}}\right)\right.\\
\left.-2\,z\,k^2+\frac{\beta}{z}-M\right]\psi'
+\frac{M_5^2\,R^2}{z^2}\psi=M_n^2\,\psi.
\end{multline}

Solutions to this equation are numerical, but fixing the boundary as $\psi(\epsilon)=\epsilon^{\Delta-S}$ and $\psi(z\to\infty)=0$ we can obtain the eigenmodes spectra for heavy quarkonium. By performing a Boguliobov transformation $\psi_n\to e^{\phi/2}\sqrt{z}\,\psi_n$ we can obtain the corresponding holographic potential. 

The decay constant spectrum calculated for vector charmonium using this model is summarized in table \ref{tab:one}. Notice the difference of almost 3$\%$ in all of the decays calculated using \eqref{decay-const-2} with the results obtained with the standard formula. This systematic error is associated with the numerical instability inherent to the derivative calculation process: since the quantities diverge so fast when they approach zero, the numerical results are sensitive to the choice of the cutoff. This problem can be observed also in the scalar decay constant, where the AdS warp factor ($\sim 1/z$) goes cubic, causing a convergence problem at the boundary located near to zero. These kinds of problems can be avoided with \eqref{decay-const-2} since all of the divergent powers of $z$ are canceled when the warp factor is considered to be AdS. 

\begin{center}
\begin{table}[h]
    \begin{tabular}{||c||c|c||c|c||}
    \multicolumn{5}{||c||}{\textbf{Charmonium decay constant spectrum}}\\
    \hline
    \hline
    \textbf{$n$}     & \textbf{$f_\text{PDG}$ (MeV) } \cite{PhysRevD.98.030001} & \textbf{$f_\text{BF}$ (MeV)\cite{Braga:2019yeh}} & \textbf{$f_{\text{Ours}}$ (MeV)}&  \textbf{$\Delta\,f$}  \\
    \hline
    \hline
       1 & $416\pm5.25$  & $399$ & $410.5$  & $2.9\%$\\
       2 & $296.08\pm2.51$& $255$ & $262.6$  & $3.0\%$ \\
       3 & $187.13\pm7.61$ & $198$ & $204.1$  & $3.0\%$\\
       4 & $160.78\pm9.70$& $169$ & $174.1$  & $3.0 \%$\\
    \end{tabular}
    \caption{Comparative of the calculated decay constants using the Braga-Ferreira proposal \cite{Braga:2019yeh} with the formula \eqref{decay-const-2}. The first column shows the experimental results obtained from \cite{PhysRevD.98.030001}.  The last column is the relative error ($\Delta\, f$) between both models. Notice that same value of the error, $3\%$, tells that the differences are systematic, i.e., related to the numerical implementation in both cases. }
    \label{tab:one}
\end{table}    
\end{center} 

The strongest feature of the formula \eqref{decay-const-2} is at the numerical level: once we know the eigensolution, we can evaluate the Taylor expansion and the quotient with $z^{\Delta+S}$ and read off directly the decay constant up to normalization factors fixed by the large $N_c$ behavior at $q^2\to \infty$. In this case, we obtain the decay constant spectrum for the vector charmonia states, exposed in the third column of table \ref{tab:one}.

\subsection{Li and Huang Dilaton}
Li and Huang propose a dilaton field to address the chiral symmetry breaking in the context of a two flavor dynamical AdS/QCD model \cite{Li:2013oda}. In this approach, authors study the effect of a dilaton field

\begin{equation}
\Phi(z)=\mu^2\,z^2\,\tanh\left(\mu^2\,z^2\right),    
\end{equation}

onto the chiral expectation value, modeled by a scalar field $\chi(z)$, and the geometry background, by solving the dynamically the graviton-dilaton-scalar system. 

The warp factor $A(z)$ is constrained to be asymptotically AdS and is calculated from  (see eqn. (4.16) in \cite{Li:2013oda})

\begin{equation}
-A''+A'^2+\frac{2}{3}\phi''-\frac{4}{3}\,A'\,\Phi'-\frac{\lambda}{6}\,e^{\Phi}\chi'^2=0,    
\end{equation}

\noindent where $\lambda$ is a constant related to 5-dimensional  gravitational constant $G_5$, the  number of colors $N_c$ and flavors $N_f$.

The chiral field $\chi(z)$ is constructed from the dilaton as 

\begin{equation}
\chi'(z)=\sqrt{\frac{8}{\lambda}}\,\mu\,e^{-\frac{\Phi}{2}}\left[1+C_1\,e^{-\Phi}+C_2\,e^{-2\,\Phi}-\frac{1}{2}e^{-3\,\Phi}\right],    
\end{equation}

\noindent, where $\mu$ is an energy constant; $C_1$ and $C_2$, are constants defined in terms of the quark mass and the chiral condensate. In this sense the LH model captures the chiral symmetry breaking phenomenology. 

The hadronic part is given by the standard SWM actions exposed in \cite{Karch:2006pv} and \cite{Colangelo:2008us}, but evaluated with the dynamical background.

To follow the same procedure as we did before, we need to solve numerically the equation of motion for light vector mesons, that in this model reads 

\begin{equation}
\partial_z\left[e^{A(z)-\Phi(z)}\,\psi'_n(z,q)\right]+\left(-q^2\right)e^{A(z)-\Phi(z)}\,\psi_n(z,q)=0.  
\end{equation}

Confinement in this model is realized by the holographic potential obtained by performing a Boguliobov   transformation $\psi_n\to e^{\frac{1}{2}(\Phi-A)}\psi_n$. Figure \ref{fig:one} shows a plot of this potential for the case of $\rho$ mesons. 

In this model, decay constants are calculated using the expression (see eqn. 5.2 in \cite{Li:2013oda}):

\begin{equation}
F_n^2=\frac{N_f}{g_5^2\,N_c}\left[e^{A-\Phi}\partial_z\,\psi_n\right]_{z\to 0}, 
\end{equation}

\noindent where $N_f=2$, $N_c=3$ and $g_5^2=\frac{12 \,\pi^2}{N_c}$.

Alternatively, decay constants in this model can be calculated with the master formula \eqref{decay-const-2}. The only necessary ingredient is the numerical behavior of $\psi_n(z)/z^{\Delta-S}$ in the limit $z\to 0$.

\begin{center}
\begin{table}[h]
    \begin{tabular}{||c||c||c||c||}
    \multicolumn{4}{||c||}{\textbf{$\rho$ meson decay constant}}\\
    \hline
    \hline 
        \textbf{State} & \textbf{$f_\text{PDG}$  (MeV)} \cite{PhysRevD.98.030001} & \textbf{$f_\text{LH}$ (MeV)} & \textbf{$f_\text{Ours}$ (MeV)} \\
        \hline
        \hline 
        $\rho(770)$ &$414.12\pm0.85$ &$299$ & $297.6$ \\
        
    \end{tabular}
    \caption{Numerical results comparing the formula \eqref{decay-const-2} with the LH model \cite{Li:2013oda}. Notice the difference with the experimental value constructed from PDG. Again the numerical differences can be related to the (systematic) numerical error in the calculation.}
    \label{tab:two}
\end{table}    
\end{center}

\begin{center}
\begin{table}[b]
    \begin{tabular}{||c||c||c||c||}
    \multicolumn{4}{||c||}{\textbf{$\rho$ meson decay constant}}\\
    \hline
    \hline 
        \textbf{State} & \textbf{$f_\text{PDG}$  (MeV)} \cite{PhysRevD.98.030001} & \textbf{$f_\text{LH}$ (MeV)} & \textbf{$f_\text{Ours}$ (MeV)} \\
        \hline
        \hline 
        $\rho(770)$ &$414.12\pm0.85$ &$299$ & $297.6$ \\
        
    \end{tabular}
    \caption{Numerical results comparing the formula \eqref{decay-const-2} with the LH model \cite{Li:2013oda}. Notice the difference with the experimental value constructed from PDG. Again the numerical differences can be related to the (systematic) numerical error in the calculation.}
    \label{tab:two}
\end{table}    
\end{center}

\begin{center}
 \begin{figure*}
\begin{tabular}{c c}
\includegraphics[width= 3.3 in]{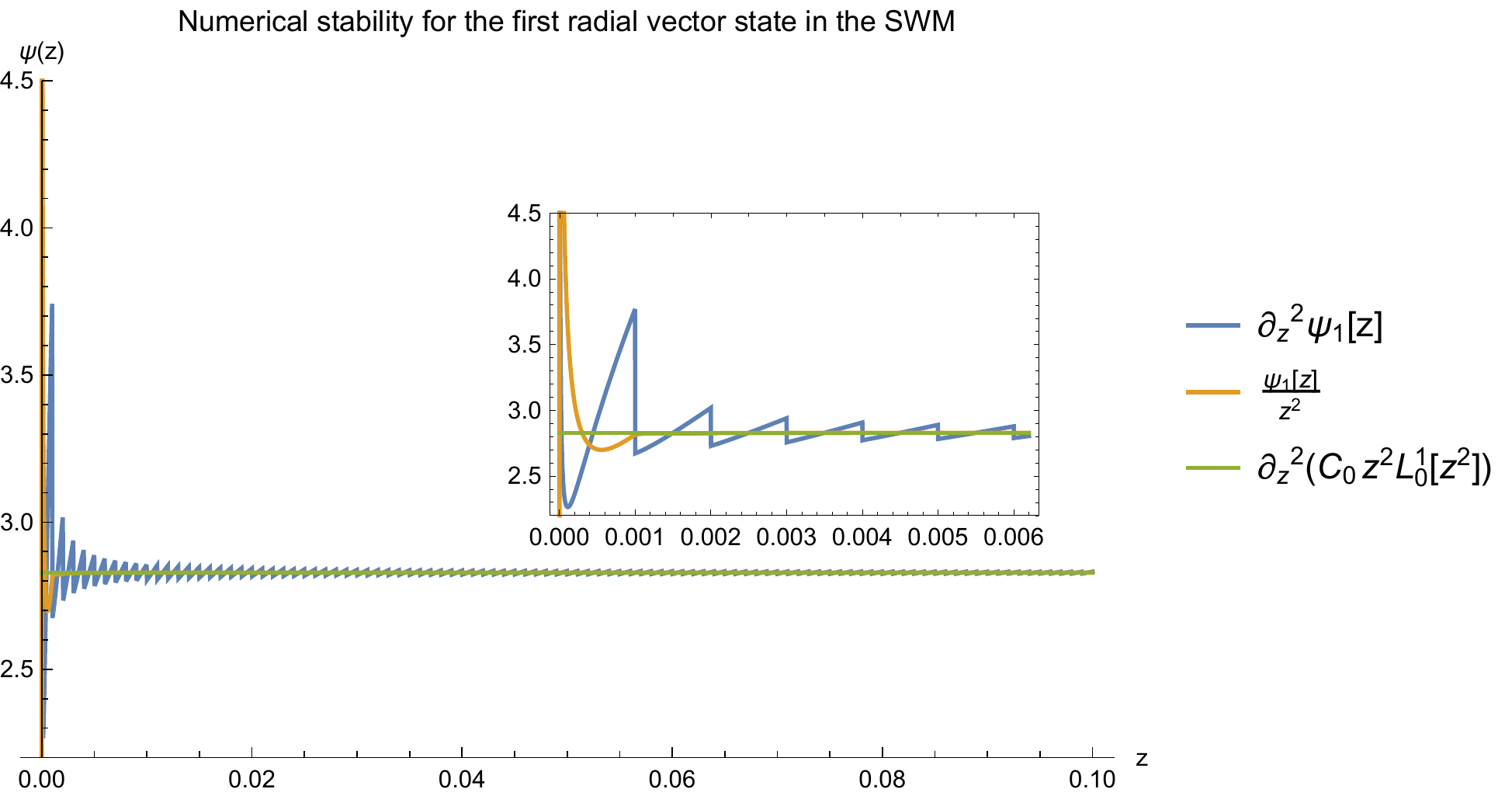}
\includegraphics[width= 3.3 in]{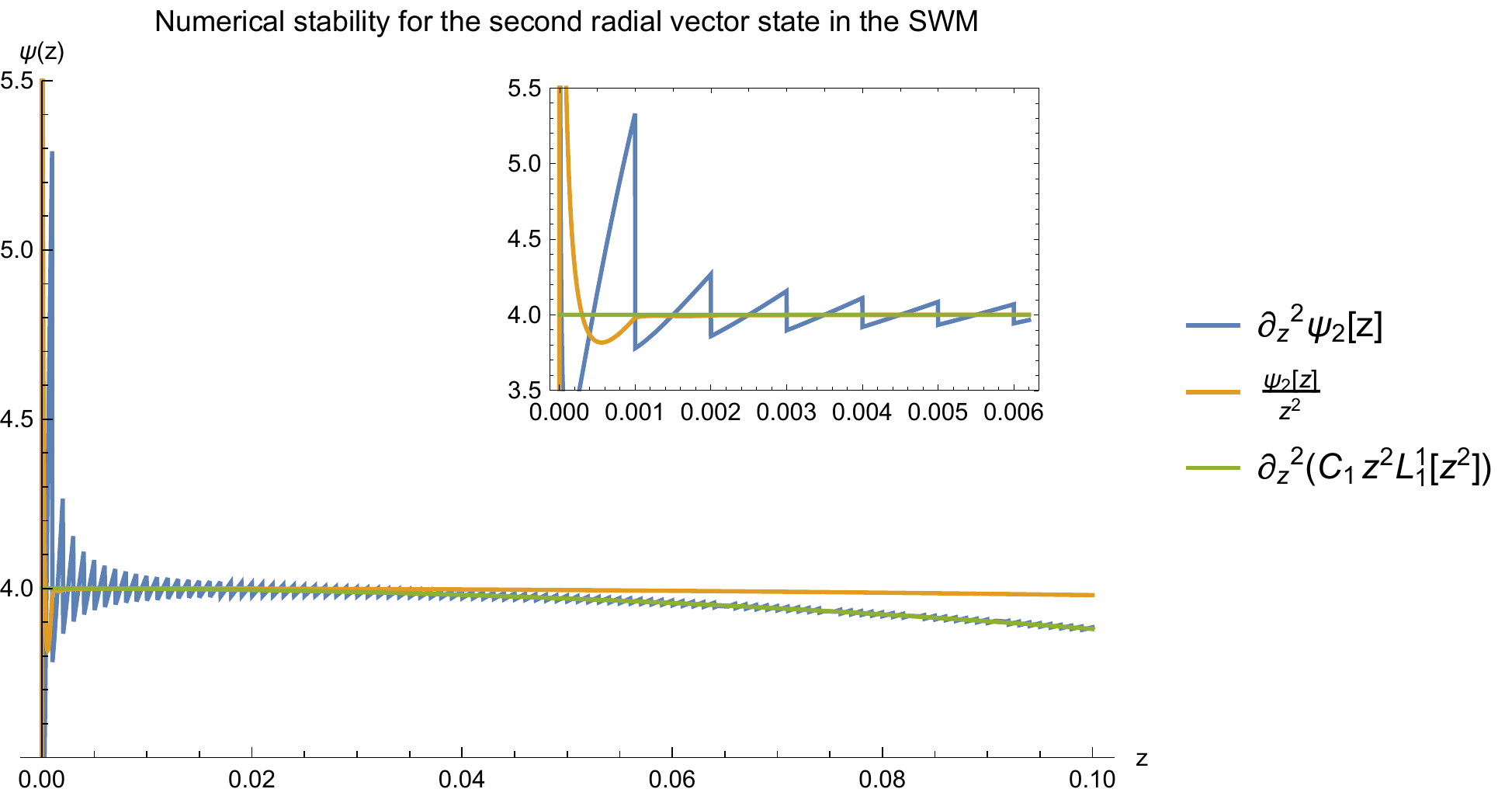}\\
\includegraphics[width= 3.3 in]{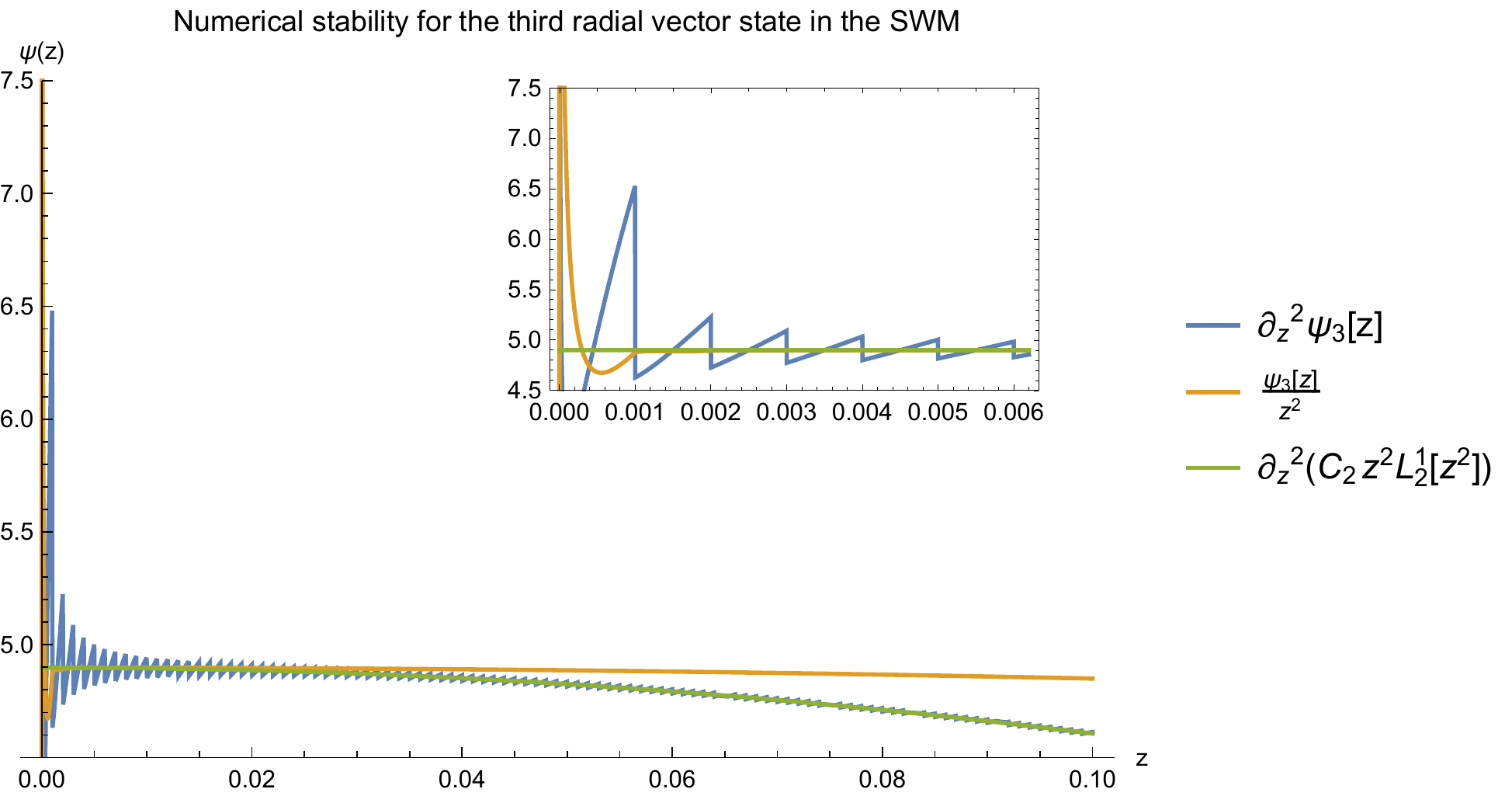}
\includegraphics[width= 3.3 in]{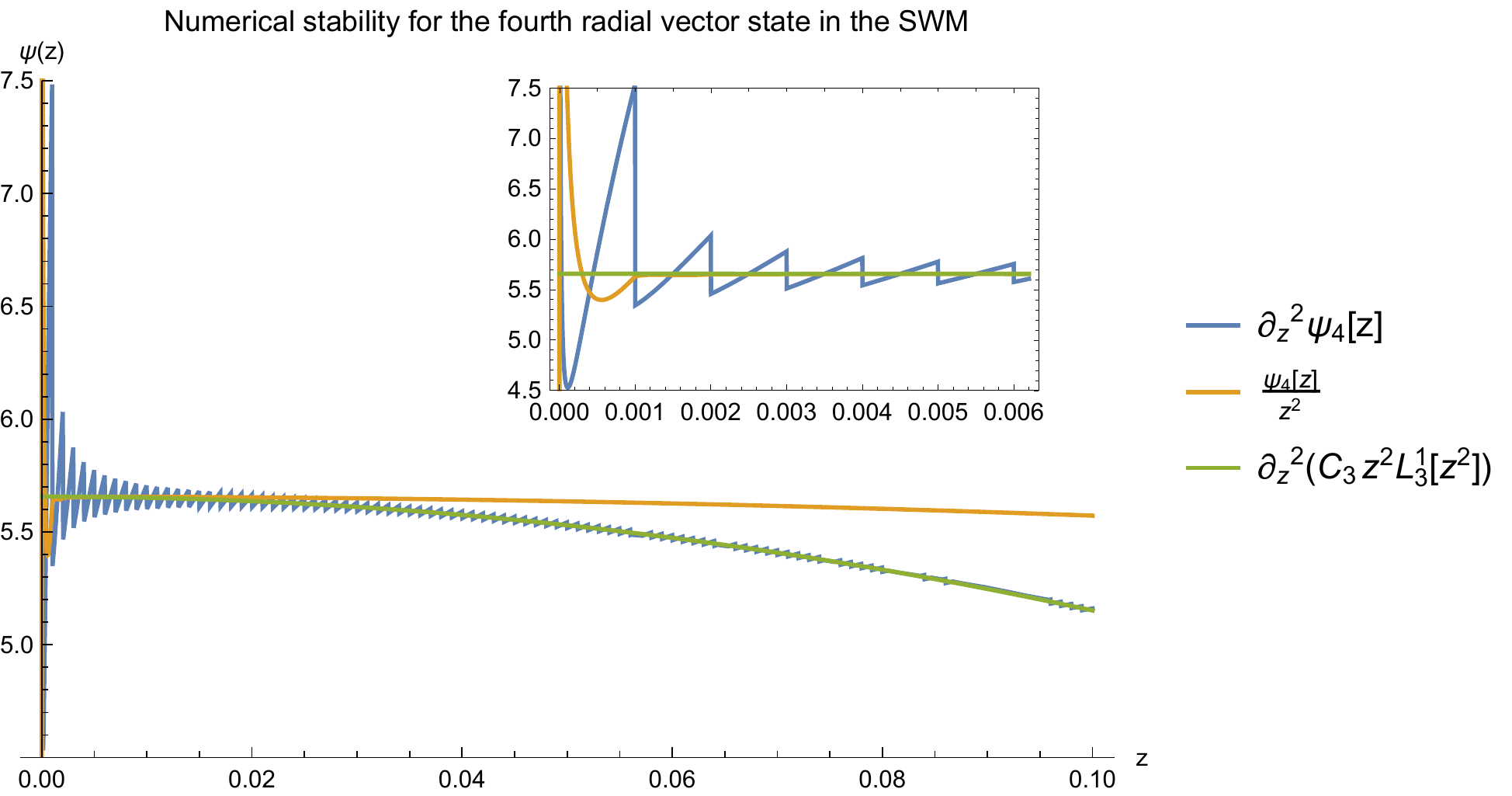} 
\end{tabular}
\caption{The low $z$ behavior of the second derivative for the first four vector states in the soft wall model, compared with corresponding $\psi_n/z^{\Delta-S}$ is plotted. Each panel has the numerical second derivative, the analytical second derivative and our proposal for the decay plotted in a $z$ interval from $0$ up to $0.1$. The small window inside each panel is a zoom of the depicted situation in the region of numerical sensitivity. Observe how in the region before $z=10^{-3}$, numerical estimations do not converge. The first one stabilizing is the $\psi/z^{2}$ near to $z=10^{-3}$. Thus any calculation before this point has no numerical reliability.}
\label{fig:two}
\end{figure*}   
\end{center} 

Numerical results for the calculation of the decay constant for $\rho(770)$ are shown in table \ref{tab:two}. Notice that there is also a systematic error present in the calculation of the decay since the instabilities associated with the numerical derivative. As in the case of the BF model, these dynamical-type models are sensitive to the behavior of the derivatives and divergent terms near to the boundary. Another ingredient here is the AdS-like behavior of the metric. Since the warp factor is not Poincare, we could be tempted to say that the master formula \eqref{decay-const-2} is not valid. But, as a manner of demonstration, just consider that any AdS-like warp factor can be written in a generic form as

\begin{equation}
A(z)=\log\frac{R}{z}+F(z),    
\end{equation}

\noindent where the deformation $F(z)$ should be a well-behaved function of the holographic coordinate $z$, that at the boundary is zero, i.e., if $z\to 0$ then $F(z)\to 0$. If this holds, then the cancellation of powers of $z$ in the decay constant expression is valid, since $e^{F(z)}$ will contribute with terms  $\mathcal{O}(z^n)$, with $n$ positive, implying that the deformation does not affect the holographic decay constant.

This simple discussion also shows the limitation if the master formula \eqref{decay-const-2}: if the geometry is allowed to be non-AdS-like, the exact cancellation of powers of $z$ is not achieved. 

\section{How to choose a proper boundary}

\begin{center}
\begin{figure*}
\begin{tabular}{c c}
    \includegraphics[width= 3.3 in]{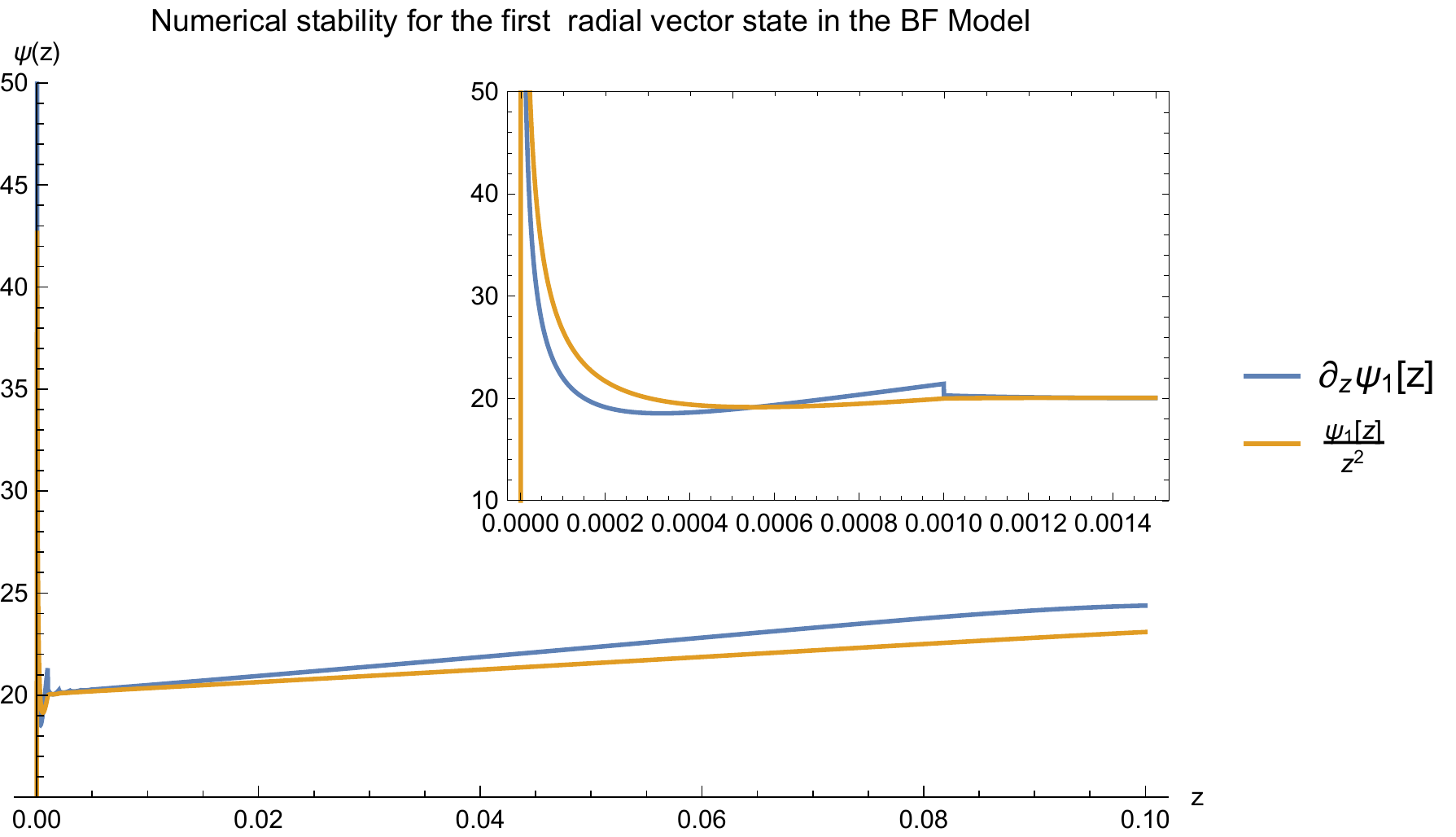}
\includegraphics[width= 3.3 in]{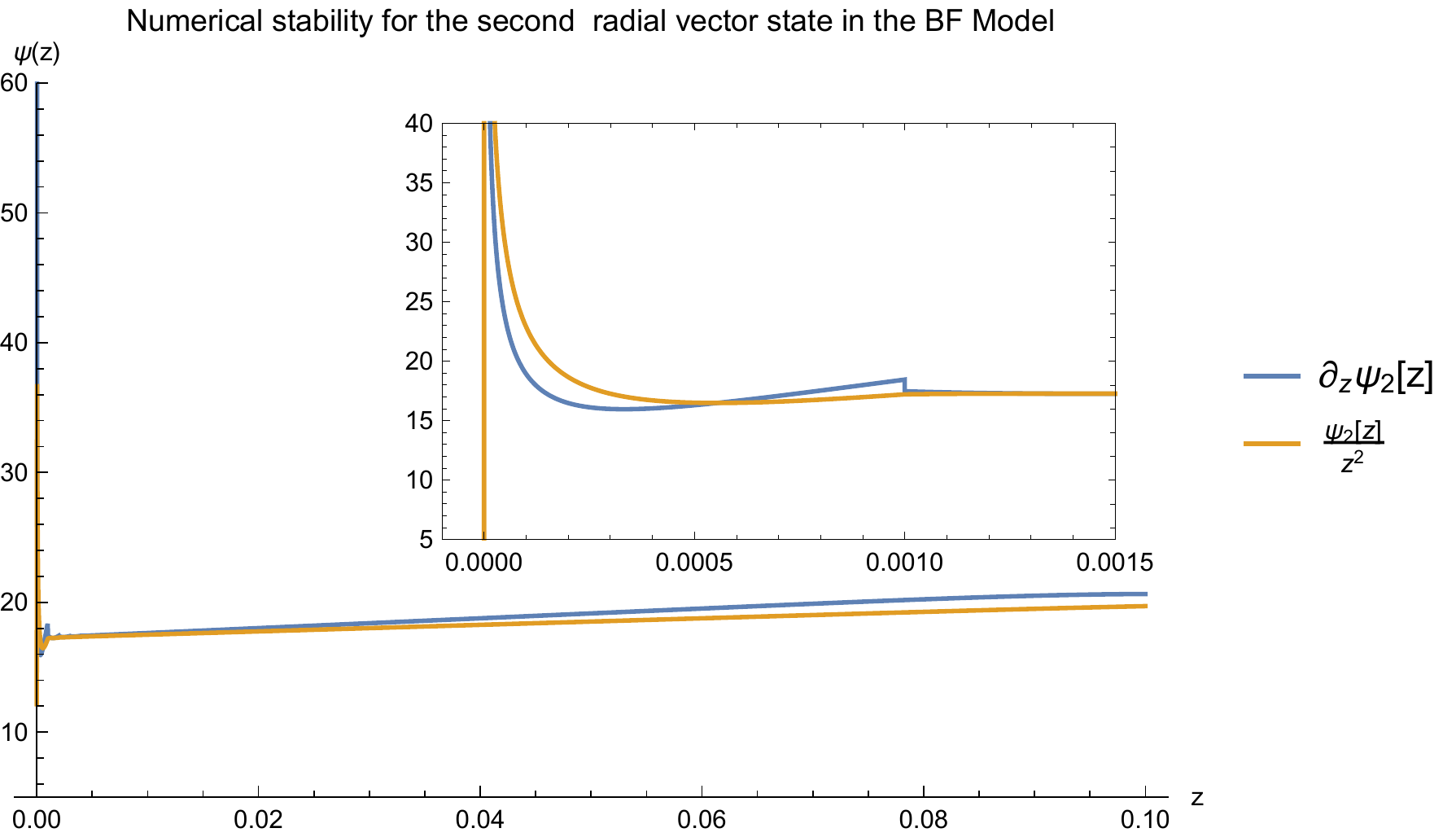}\\
\includegraphics[width= 3.3 in]{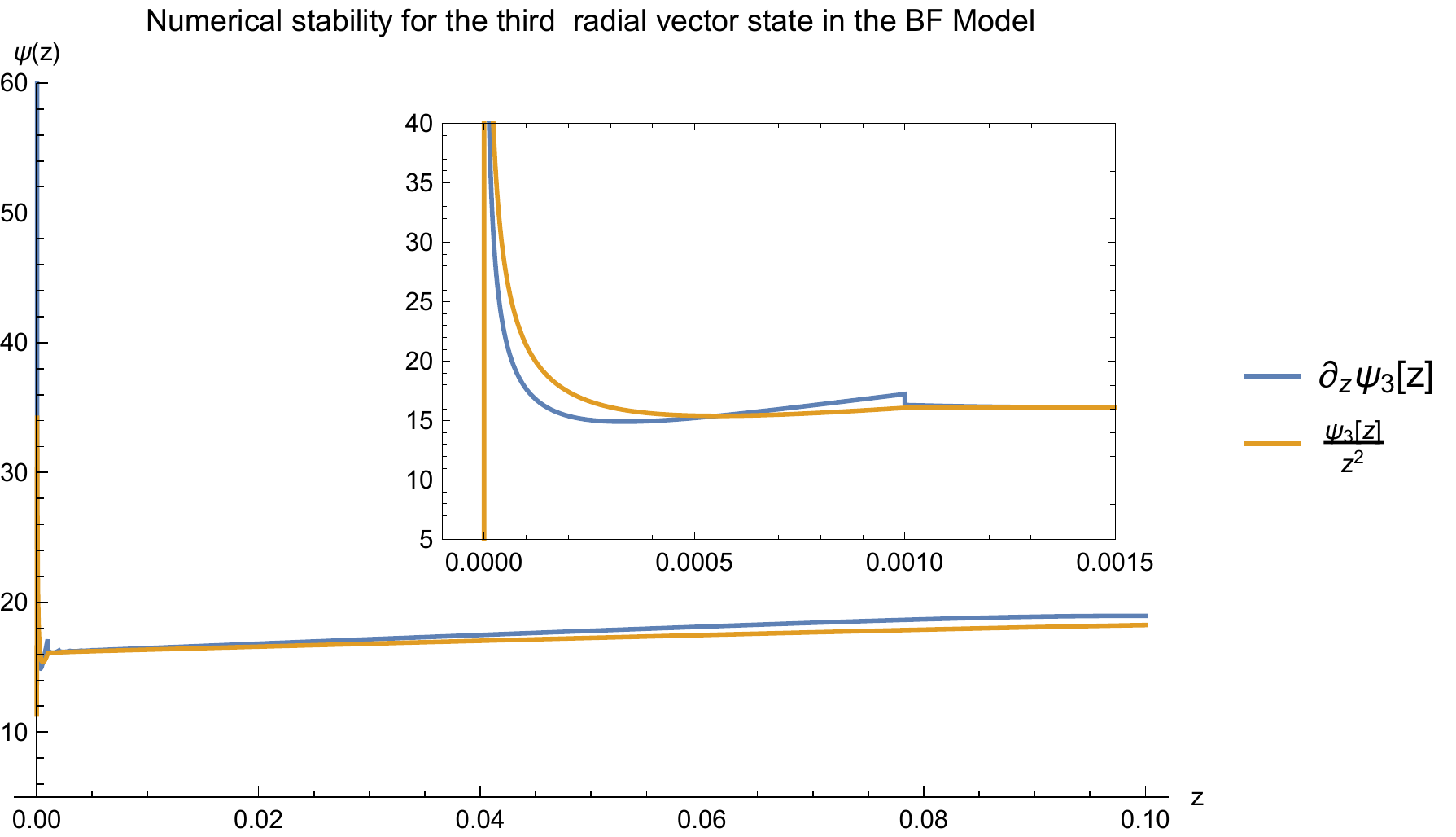}
\includegraphics[width= 3.3
in]{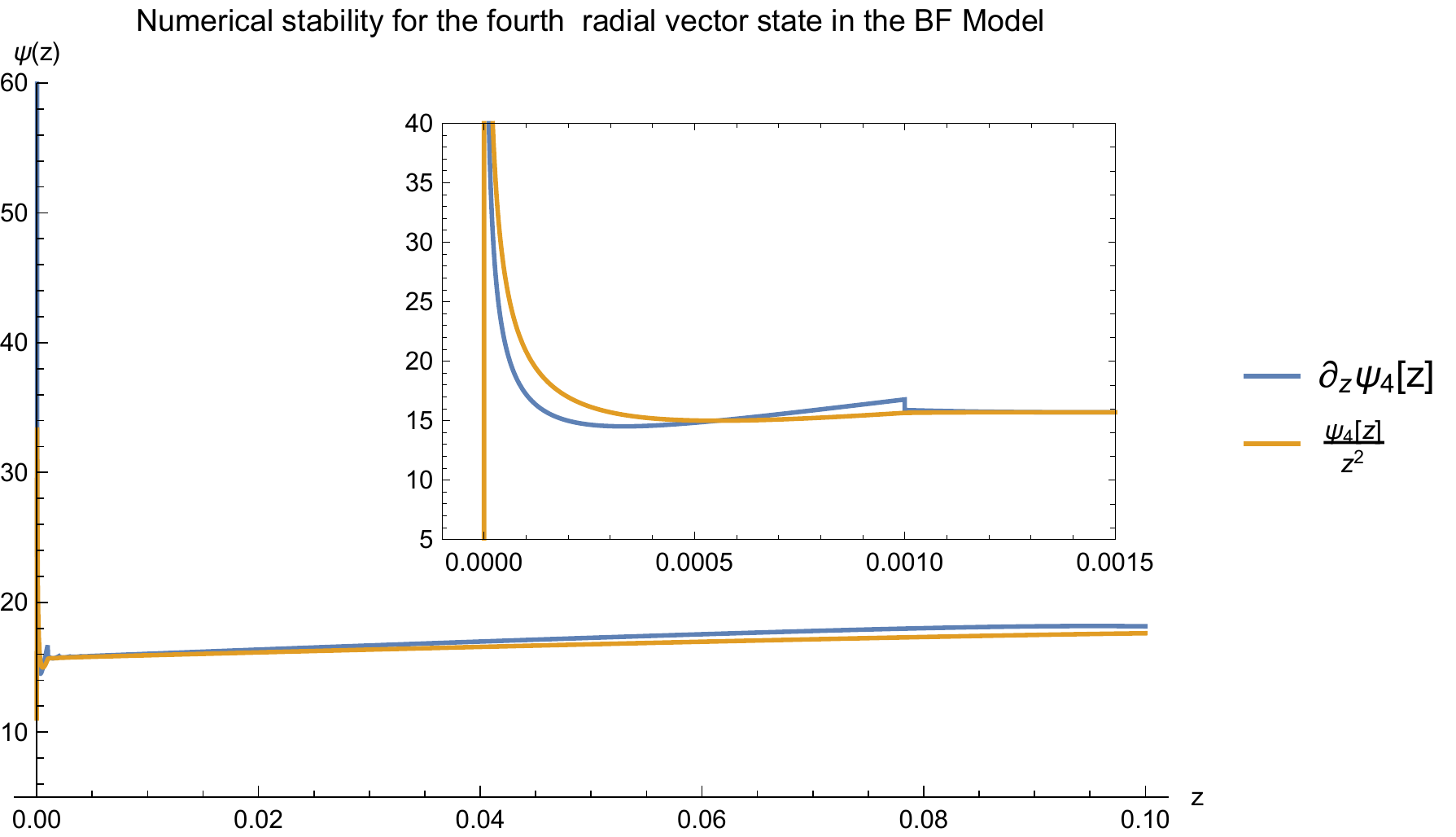} 
\end{tabular}
\caption{The low $z$ behavior of the first derivative for the first four vector states in the soft wall model, compared with corresponding $\psi_n/z^{\Delta-S}$ plotted. Unlike the soft wall model case, the BF model is not analytical therefore we cannot compare against the analytical solution. Thus, the criterion, in this case, is set by the convergence of both solutions in the low $z$ limit. Again, each panel ranges from $z=0$ up to $z=0.01$. Each panel displays a zoom-in of the convergence region also. Notice that, in most of the cases, $\psi/z^2$ power expansion converges faster than the numerical derivative calculation. Notice that before $z=5.0\times10^{-4}$, decay constant numerical calculations are not reliable.}
\label{fig:three}
\end{figure*}    
\end{center}

Summarizing, the most important conclusion we made up to this point by comparing the calculation process of several AdS/QCD models is a strong dependence of the decay with the boundary locus. At first sight, this $3\%$ level of confidence could be meaningless, but numerically it is important. The conformal boundary used to compute eigenstates from the usual AdS/QCD  potentials depicted in figure \ref{fig:one} is quite different from the one used in the expression \eqref{decay-cons-1}. From the Sturm-Liouville point of view, the eigenvalue spectrum in these particular cases is always defined by Dirichlet boundary conditions that are not so sensitive to numerical boundaries chosen in the range of $\epsilon\rightarrow10^{-7}$ or $10^{-2}$. The calculation of the decay constants is controlled by Neumann boundary conditions, which numerically are sensitive to small numerical differences due to the derivative. In this sense, numerical calculations at the numerical boundary used for the eigenvalue spectrum are unstable. These calculations tend to stabilize at bigger values that the eigenvalue boundary. This point is illustrated in figures \ref{fig:two} and \ref{fig:three} for the first vector states calculated in the soft wall model and Braga--Ferreira model.  

This analysis suggests that the choice of the conformal boundary for numerical calculations is not a trivial task. In fact, one should first plot the (first or second) derivative of the mesonic eigenstates and check wherein the low $z$ region the decays stabilize. This low $z$ issue is more perceptible in the high excitation number part of the decay constants spectrum. Fortunately, most of the mesonic spectra have not so high measured excited states.  Therefore, in this particular sense, any AdS/QCD model that tries to calculate the decay constants have to fix properly the boundary first. This is similar to the UV cutoff AdS/QCD models \cite{Braga:2015jca,Braga:2017bml} where the boundary has a finite (no-null) boundary.

\section{Conclusions and Final Comments}\label{conclu}

In this work we have developed an alternative form to calculate decay constants, avoiding the numerical interference attached to derivatives and divergent quantities near to the conformal boundary. The formula \eqref{decay-const-2} connects the decay constant of a mesonic state with the lowest coefficient $\mathcal{C}_{\Delta-S}$ of the Taylor expansion for the eigenmode $\psi_n(z,q)$, evading the numerical issues exposed above. 

From the phenomenological point of view, we prove that decay constants and AdS modes to zero are related when $z$ go, which is similar to the Van Royen - Weisskopf formula, so equation \eqref{decay-cons-3} could be considered as a holographic realization of this famous expression. The constant $\mathcal{C}_{\Delta-S}$ appearing in the Taylor expansion carries all of the information about the eigenmode $\psi_n$ at the boundary. This information, as was proved by \cite{Karch:2006pv}, is one of the two things relevant to know the decay constant.  

The other is the proper value of $\mathcal{K}$. that comes from the matching of the holographic 2-point function with the corresponding large $N_c$-QCD 2-point function, both at the large $q^2$ limit \cite{Afonin:2004yb,Hong:2004sa}. If the dilaton field vanishes at the boundary, as in the case of the soft wall model, constant $\mathcal{K}$ is fixed to be $g_5^2=4\,\pi^2$ for a theory with $N_f=2$ and $N_c=3$. But, on the other hand, if the dilaton does not vanish, then the constant $\mathcal{K}$ should include the information about the value of the dilaton at the boundary. If we perform the large $Q^2=-q^2$ expansion of the holographic vector 2-point function for an AdS-like bottom-up model with dilaton, we can arrive at the leading order expression 

\begin{equation}
\Pi_{V,\text{hol}}(Q^2)=-\frac{e^{-2\,\Phi_0}}{2\,\mathcal{K}}\,\log Q^2;    
\end{equation}

\noindent such that, if we compare with the corresponding large $N_c$ QCD object, also at leading order, 

\begin{equation}
\Pi_V(Q^2)=-\frac{N_c}{24\,\pi^2}\log Q^2,    
\end{equation}

\noindent therefore, we conclude that the normalization constant $\mathcal{K}$ should be fixed to be keeping in mind the value of the dilaton at the boundary, i.e., $\Phi_0$. This fixing process of $\mathcal{K}$ seems meaningless since it does not affect directly the confinement properties observed in the scalar and vector mesons spectra. But in the case of the pseudoscalar and axial mesons, $\mathcal{K}$ plays an important role since it appears directly in the respective equations of motion for their eigenfunctions \cite{Colangelo:2008us}. But, since most of the deformed dilatons introduced after soft wall model are constrained to be vanishing at the boundary and quadratic at high $z$ (see for example \cite{Batell:2008zm,He:2013qq,FolcoCapossoli:2018phk}), the effect of the dilaton in $\mathcal{K}$ was not explored. But, this feature opens the door to other possibilities to try to fit decay constants by playing with the low $z$ behavior of the dilaton. This is an interesting topic that will be addressed in future works.

\begin{acknowledgments}
We wish to acknowledge Prof. Dr. D. Li, at Jinan University in China, for all of the useful discussions. We also acknowledge the financial support of FONDECYT (Chile) under Grants No. 1180753 (A. V.) and No. 3180592 (M. A. M. C.).
\end{acknowledgments}

\end{document}